\documentclass{pasj00}
\SetRunningHead{X-ray Groups in zCOSMOS}{Tanaka et al.}

\usepackage{times}
\usepackage{ulem}

\begin{document}

\title{X-ray Groups of Galaxies at $0.5<z<1$ in zCOSMOS:\\
Increased AGN Activities in High Redshift Groups}
\author{
M.~\textsc{Tanaka}\altaffilmark{1},
A.~\textsc{Finoguenov}\altaffilmark{2},
S.~J.~\textsc{Lilly}\altaffilmark{3},
M.~\textsc{Bolzonella}\altaffilmark{11},
C.~M.~\textsc{Carollo}\altaffilmark{3},
T.~\textsc{Contini}\altaffilmark{4,5},
A.~\textsc{Iovino}\altaffilmark{6},
J.-P.~\textsc{Kneib}\altaffilmark{7},
F.~\textsc{Lamareille}\altaffilmark{4,5}
O.~\textsc{Le Fevre}\altaffilmark{7},
V.~\textsc{Mainieri}\altaffilmark{8},
V.~\textsc{Presotto}\altaffilmark{6},
A.~\textsc{Renzini}\altaffilmark{9},
M.~\textsc{Scodeggio}\altaffilmark{10},
J.~D.~\textsc{Silverman}\altaffilmark{1},
G.~\textsc{Zamorani}\altaffilmark{11},
S.~\textsc{Bardelli}\altaffilmark{11},
A.~\textsc{Bongiorno}\altaffilmark{2},
K.~\textsc{Caputi}\altaffilmark{12},
O.~\textsc{Cucciati}\altaffilmark{6},
S.~\textsc{de la Torre}\altaffilmark{12},
L.~\textsc{de Ravel}\altaffilmark{12},
P.~\textsc{Franzetti}\altaffilmark{10},
B.~\textsc{Garilli}\altaffilmark{10},
P.~\textsc{Kampczyk}\altaffilmark{3},
C.~\textsc{Knobel}\altaffilmark{3},
K.~\textsc{Kova\u{c}}\altaffilmark{3,13},
J.-F.~\textsc{Le Borgne}\altaffilmark{4,5},
V.~\textsc{Le Brun}\altaffilmark{7},
C.~\textsc{L\'opez-Sanjuan}\altaffilmark{7},
C.~\textsc{Maier}\altaffilmark{3},
M.~\textsc{Mignoli}\altaffilmark{11},
R.~\textsc{Pello}\altaffilmark{4,5},
Y.~\textsc{Peng}\altaffilmark{4,5}
E.~\textsc{Perez Montero}\altaffilmark{4,5,14},
L.~\textsc{Tasca}\altaffilmark{7},
L.~\textsc{Tresse}\altaffilmark{7},
D.~\textsc{Vergani}\altaffilmark{11},
E.~\textsc{Zucca}\altaffilmark{11},
L.~\textsc{Barnes}\altaffilmark{3},
R.~\textsc{Bordoloi}\altaffilmark{3},
A.~\textsc{Cappi}\altaffilmark{11},
A.~\textsc{Cimatti}\altaffilmark{15},
G.~\textsc{Coppa}\altaffilmark{2},
A.~M.~\textsc{Koekemoer}\altaffilmark{16},
H.~J.~\textsc{McCracken}\altaffilmark{17},
M.~\textsc{Moresco}\altaffilmark{15},
P.~\textsc{Nair}\altaffilmark{11},
P.~\textsc{Oesch}\altaffilmark{3},
L.~\textsc{Pozzetti}\altaffilmark{11},
N.~\textsc{Welikala}\altaffilmark{18}
}
%\thanks{Last update: January 19, 2007}

\altaffiltext{1}{Institute for the Physics and Mathematics of the Universe, The University of Tokyo,  5-1-5 Kashiwanoha, Kashiwa-shi, Chiba 277-8583, Japan}
\altaffiltext{2}{Max-Planck Institut f\"{u}r extraterrestrische Physik, Giessenbachstrasse, D-85748 Garching bei M\"{u}nchen, Germany}
\altaffiltext{3}{Institute of Astronomy, ETH Z\"{u}rich, Z\"{u}rich 8093, Switzerland}
\altaffiltext{4}{Institut de Recherche en Astrophysique et Plan\'{e}tologie, CNRS, 14, avenue Edouard Belin, F-31400 Toulouse, France}
\altaffiltext{5}{IRAP, Universit\'{e} de Toulouse, UPS-OMP, Toulouse, France}
\altaffiltext{6}{INAF Osservatorio Astronomico di Brera, Milan, Italy}
\altaffiltext{7}{Laboratoire d'Astrophyque de Marseille, CNRS/Aix-Marseille Universit\'{e}, 38 rue Fr\'{e}d\'{e}ric Joliot-Curie, 13388, Marseille cedex 13, France}
\altaffiltext{8}{European Southern Observatory, Karl-Schwarzschild-Str. 2, D-85748 Garching bei M\"{u}nchen, Germany}
\altaffiltext{9}{Dipartimento di Astronomia, Universita di Padova, Pavoda, Italy}
\altaffiltext{10}{INAF - IASF Milano, Milan, Italy}
\altaffiltext{11}{INAF Osservatorio Astronomico di Bologna, Bologna, Italy}
\altaffiltext{12}{Institute for Astronomy, University of Edinburgh, Royal Observatory, Blackford Hill, Edinburgh EH9 3HJ, UK}
\altaffiltext{13}{Max-Planck Institut f\"{u}r Astrophysik, D-85748, Garching bei M\"{u}nchen, Germany}
\altaffiltext{14}{Instituto de Astrof\'{i}sica de Andaluc\'{i}a, CSIC, Apartado de correos 3004, 18080 Granada, Spain}
\altaffiltext{15}{Dipartimento di Astronomia, Universit\`{a} degli Studi di Bologna, Bologna, Italy}
\altaffiltext{16}{Space Telescope Science Institute, Baltimore, Maryland 21218, USA}
\altaffiltext{17}{Institut d'Astrophysique de Paris, UMR7095 CNRS, Universit\'{e} Pierre \& Marie Curie, 75014 Paris, France}
\altaffiltext{18}{Insitut d'Astrophysique Spatiale, CNRS \& Universit de Paris Sud-XI, 91405 Orsay Cedex, France}

%\email{masayuki.tanaka@ipmu.jp}
\KeyWords{surveys, galaxies: evolution, galaxies: fundamental parameters, galaxies: clusters: general}

\maketitle

%-------------------------------------------------------------------------------
%-------------------------------------------------------------------------------
\begin{abstract}
We present a photometric and spectroscopic study of galaxies
at $0.5<z<1$ as a function of environment based on data from the zCOSMOS survey.
There is a fair amount of evidence that galaxy properties
depend on mass of groups and clusters, in the sense that quiescent
galaxies prefer more massive systems.  We base our analysis on a mass-selected
environment using X-ray groups of galaxies and define the group membership 
using a large number of spectroscopic redshifts from zCOSMOS.
We show that the fraction of red galaxies is higher in groups than in the field
at all redshifts probed in our study.  Interestingly, the fraction
of {\sc [oii]} emitters on the red sequence increases at higher redshifts
in groups, while the fraction does not strongly evolve in the field.
This is due to increased dusty star formation activities and/or 
increased activities of active galactic nuclei (AGNs) in high redshift groups. 
We study these possibilities using the 30-band photometry and X-ray data.
We find that the stellar population of the red {\sc [oii]} emitters
in groups is old and there is no clear hint of dusty star formation
activities in those galaxies.  The observed increase of red {\sc [oii]} emitters
in groups is likely due to increased AGN activities.
However, our overall statistics is poor and any firm conclusions
need to be drawn from a larger statistical sample of $z\sim1$ groups.
\end{abstract}

%-------------------------------------------------------------------------------
\section{Introduction}

The matter distribution in the early universe is nearly uniform,
but not completely so, and small density fluctuations grow with time
through gravitational forces.
Eventually, matter becomes dense enough to initiate star formation
and galaxies form in high density peaks of the density fluctuations.
Galaxies grow progressively more massive by accreting material
from the surroundings and by merging with other galaxies.
The cosmic large-scale structure, in which galaxies are embedded,
also develops with time.  Later, clusters of galaxies
form at the nodes of filaments, where galaxies can be quenched by
gravitational and gas-dynamical effects.
The evolution of galaxies and large-scale structure goes
in tandem and galaxies eventually acquire the properties that
we observe today.
This current framework of galaxy formation and evolution indicates
that the formation and evolution of galaxies are driven by statistical events
(e.g.,  they form in density fluctuations and grow by mergers)
and galaxies are statistical objects in nature.  They do not form at the same
time and they do not evolve in the same way.  Only by statistical analyses
can we study the physics of galaxy formation and evolution.

This fundamental principle has motivated a number of galaxy surveys.
Imaging surveys deliver limited information about galaxy properties
because one needs precise distances to galaxies in order to translate
observed quantities into physical quantities.
For this reason, several large spectroscopic surveys have been
carried out to date and they have brought new insights into galaxy
evolution over cosmological time scales.

The CfA redshift survey carried out the first systematic spectroscopic
survey of the local universe in the late-70's \citep{geller89}.
They measured redshifts of nearby galaxies and revealed the
cosmic large-scale structure in the local universe, making
a major progress in our understanding of galaxy distribution
in the universe.
Following the CfA redshift survey, the Las Campanas redshift survey
\citep{shectman96} mapped out the galaxy distribution out to larger distances,
and the Canada-France redshift survey \citep{lilly95} reached out to $z=1$.
The Sloan Digital Sky Survey (SDSS; \cite{york00}) and the 2 degree field
redshift survey (2dF; \cite{colless03}) surpassed the previous surveys with
much improved statistics.  In particular, 
SDSS imaged a quarter of the sky in five photometric bands \citep{fukugita96,doi10}
and measured more than 2 million redshifts with unprecedented precision \citep{aihara11}.
The SDSS dramatically refined our view of the local universe.
In parallel to these surveys of the local universe, large spectroscopic
surveys with 8m telescopes
such as DEEP2 \citep{davis03}, VIMOS VLT Deep Survey \citep{lefevre05},
and zCOSMOS \citep{lilly07}
peered deep into the universe reaching $z>1$.
All of these surveys enabled statistical analyses of galaxy
populations over a large redshift range, bringing in
new insights into the evolution of galaxies.
This paper is in a context of such statistical galaxy studies
from large spectroscopic surveys.
We will study galaxy properties out to $z=1$ using data from
zCOSMOS with a particular emphasis on the dependence of
galaxy properties on environment.

A pioneering work on this subject was made by \citet{dressler80},
who first quantified the morphology-density relation.
Following this work, many authors studied the relationship
between galaxy properties and environment
(e.g., \cite{postman84,whitmore93,balogh97,dressler97,poggianti99};
see \cite{tanaka05} for a thorough set of references
and also see below for more recent papers).
The SDSS and 2dF data sets delivered the unprecedented
statistics and we now have a fairly good understanding of 
galaxy properties in the local universe  \citep{lewis02,gomez03,blanton03,goto03,kauffmann04,tanaka04,blanton05,baldry06}.
The advent of 8m-class telescopes pushed those environment studies
to redshift of unity and beyond (
\cite{rosati99,kodama01,lubin02,demarco05,nakata05,poggianti06,stanford05,tanaka05,stanford06,demarco07,koyama07,fassbender08,lidman08,poggianti08,mei09,bauer10,rettura10,strazzullo10,tanaka10a,tanaka10b}).
However, many of these high-$z$ studies still suffer from limited statistics 
particularly in low-medium density environments and
this is the area where large spectroscopic surveys fill in
as they mainly probe such environments
\citep{cucciati06,cooper07,tasca09,bolzonella10,cucciati10,cooper10,iovino10}.

However, results from these deep surveys and their interpretations
are not always consistent.
In particular, the dependence of galaxy colors on environment
at $z\sim1$ is controversial as discussed in \citet{cooper10}.
There are a number of differences between the data sets
from different surveys and the ways the analyses are made,
which might explain the possible inconsistency.
In this paper, we study photometric and spectroscopic properties
of galaxies as a function of environment and make an attempt
to settle the issue.
A unique feature of our study is that we define mass-selected
environments using X-ray groups.
Most of the previous studies are based on environment
traced by galaxies and there is room for observational biases
to come in.  Such biases include sampling rate, redshift success rate
as functions of redshift and galaxy type, large-scale structure, etc.
For example, if a sample is biased towards star forming galaxies,
which can be the case at high redshifts in optical surveys,
the density field traced by galaxies is basically the density of
star forming galaxies, which may not represent the true density field.
X-rays, on the other hand, are free from such biases.
An extended X-ray emission is a strong signature of a dynamically
bound system.  Also, an X-ray luminosity is a good proxy
for mass of a system,
making it possible to define environments by mass.
We present a robust analysis of the dependence of galaxy properties
on environment based on stellar mass limited galaxy sample with
mass-selected environments.

The structure of this paper is as follows.
We briefly summarize the zCOSMOS survey and construct an X-ray group
catalog in Section 2.
We then study properties of the group galaxies and make comparisons to field galaxies
in Section 3.
Section 4 discusses the results and the paper is concluded in Section 5.
Unless otherwise stated, we use $\Omega_M=0.26$, $\Omega_\Lambda=0.74$, and
$\rm H_0=72\ km\ s^{-1}\ Mpc^{-1}$.
The uncertainties are given in 68\% confidence intervals.
All the magnitudes are given in the AB system.

%-------------------------------------------------------------------------------
\section{X-ray Group and Member Galaxy Catalogs}

We study spectroscopic properties of galaxies up to $z=1$
based on data from the zCOSMOS survey \citep{lilly07}.
We first briefly describe the survey and move on to construct
a catalog of galaxy groups selected from deep X-ray data.

%-------------------------------------------------------------------------------
\subsection{The zCOSMOS redshift survey}

The zCOSMOS survey is a spectroscopic survey of the COSMOS field \citep{scoville07,koekemoer07}
using the VIMOS spectrograph on the Very Large Telescope on the Cerro Paranal \citep{lefevre03}.
It is the largest program ever conducted at VLT with 600 hours of allocated time.
The survey consists of two components: zCOSMOS-bright and zCOSMOS-faint.
The former is a flux-limited survey down to $I_{AB}=22.5$ using the medium
resolution grism ($R\sim500$) with a wavelength coverage of $5550-9650\rm \AA$.
The latter is a color-selected galaxy survey aiming at $z\sim2$ galaxies
using the low-resolution blue grism with $R\sim200$ over $3700-6700\rm \AA$.
All the spectra are visually inspected by two people independently
and final redshifts and confidence flags are determined through
face-face reconciliation meetings.
In this study, we use the zCOSMOS-bright 20k data to perform statistical
analysis of galaxy groups.
For further derails of the survey, the readers are referred to
\citet{lilly07} and \citet{lilly09}.

%-------------------------------------------------------------------------------
\subsection{X-ray group catalog}
We construct an X-ray group catalog using X-ray data from XMM-Newton and
Chandra available in the COSMOS field \citep{hasinger07,elvis09}.
An early version of the group catalog was presented in \citet{finoguenov07},
in which we relied only on photometric redshifts to identify the optical
counterparts of extended X-ray emission.
We revise the catalog with
an efficient group identification algorithm demonstrated by \citet{bielby10}
and \citet{finoguenov10} to the massive COSMOS data set. 
We give the full details in Finoguenov et al. (in prep), and here we only
briefly outline our algorithm.

On the mosaic of coadded XMM and Chandra images, cluster candidates are first
identified as extended sources through a classical wavelet transform technique
with a careful removal of point sources \citep{finoguenov09}.
Next, we look for the cluster red sequence around the extended X-ray sources
using the deep optical-IR data.  For this red sequence search, we construct
model red sequence using the recipe by \citet{lidman08} and quantify
a significance of red sequence around an extended X-ray source at a given redshift
in the following manner:

\begin{itemize}
\item We extract galaxies located within 1 Mpc (physical) from the X-ray center
and have $|z_{phot}-z|<0.1$, where $z_{phot}$ is the photometric redshift
of a galaxy and $z$ is the redshift at which we want to
quantify the red sequence.
\item We count galaxies with weights according to their spatial locations
from the X-ray center and to their location on a color-magnitude diagram.
Bright red galaxies located at the center have the highest weight.
\item We compare the count with the average count and its dispersion,
which are derived by placing the apertures of the same size at random
positions in the COSMOS field, to quantify the significance of
the red sequence.
\item We repeat the above procedure at $0<z<2.5$ to identify peaks of
the red sequence signals.
\item Finally, we visually inspect all the significant peaks and assign
redshift and confidence flag to the X-ray source.
\end{itemize}

In the first procedure, we make use of the excellent photometric redshift
in the COSMOS field  \citep{ilbert09} to efficiently eliminate fore-/background contamination.
In the second step, we change the filter combinations to compute
colors with redshift.  We have to make a compromise between the depth of
the data and filter combinations to probe similar rest-frame wavelengths
at different redshifts, but we always straddle the 4000\AA\ break,
which is a sensitive feature to star formation.
To be specific, we use the following color-magnitude diagrams.

\begin{itemize}
\item $0.0<z<0.3$ : $u-r$ vs. $r$
\item $0.3<z<0.6$ : $B-i$ vs. $i$
\item $0.6<z<1.0$ : $r-z$ vs. $z$
\item $1.0<z<1.5$ : $i-K_S$ vs. $K_S$
\item $1.5<z<2.5$ : $z-3.6\mu m$ vs. $3.6\mu m$
\end{itemize}

Finally, all the significant red sequence signals are visually inspected and
the group redshifts and confidence flags are assigned.
We use the 20k redshifts from zCOSMOS \citep{lilly07}
to help identify the systems and obtain spectroscopic redshifts of them
in this final identification procedure.
Details of the confidence flags will be described in Finoguenov et al. (in prep.),
but in short, we give {\sc Flag}=1 to groups that are unambiguously confirmed
with spectroscopic redshifts
and the X-ray centroid is reliably determined by high significance X-ray fluxes.
{\sc Flag}=2 is for spectroscopically confirmed groups whose X-ray center can potentially be off
(up to 30 arcsec) due to low fluxes or to blending with other sources,
and the location of the optical counterparts were used in
the centroid determinations.
We have confirmed that these uncertain centroids do not affect our
results as we discuss below.
{\sc Flag}=3 groups are likely real groups but they are not spectroscopically
confirmed yet.
The catalog contains 215 groups
with {\sc Flag}=1, 2 or 3, of which 195 are at $z<1$.
We note that $\sim90$\% (175 out of 195) of the groups at $z<1$
are spectroscopically confirmed (i.e., {\sc Flag}=1 or 2).
We do not use those {\sc Flag}=3 groups in this work, but
we have confirmed that our results do not change if we include them.

%-------------------------------------------------------------------------------
\subsection{Group members from zCOSMOS}

We define the group membership using the X-ray group catalog
constructed above and the spectroscopic redshifts from zCOSMOS.
In this work, we apply the following selection criteria
to study the dependence of galaxy properties on environment:

\begin{enumerate}
\item We use spectroscopically confirmed groups at $0.5<z<1.0$ with
      masses between $3\times10^{13}\rm M_\odot$ and $7\times10^{13}\rm M_\odot$.
\item We define group members as galaxies with high confidence spectroscopic redshifts
      from zCOSMOS and located within $<2\sigma$ and $<R_{200}$ from the group centers,
      where $\sigma$ is line-of-sight velocity dispersion and $R_{200}$ is a virial radius of a group
      within which the mean interior density is 200 times the critical density of
      the universe at the group redshift.
\end{enumerate}

\begin{figure}
  \begin{center}
    \FigureFile(90mm,1mm){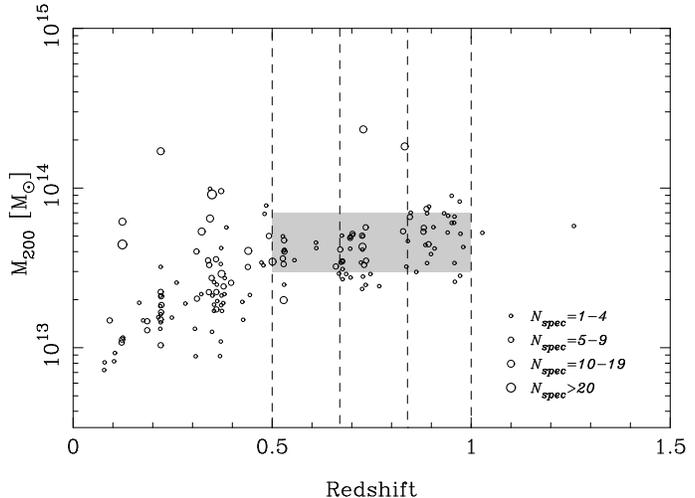}
  \end{center}
  \caption{
    $M_{200}$ of groups measured from X-rays plotted against redshift.
    We use groups with high confidence flags only.
    We make three redshift bins ($0.5<z<0.67$, $0.67<z<0.84$, and $0.84<z<1.00$)
    as indicated by the vertical dashed lines, where we have the {\sc [oii]} line
    in the spectral wavelength coverage and we are not strongly affected by
    fringes in the spectra.  In order to minimize the group mass dependence
    of galaxy properties, we use a narrow mass range as shown by the shade
    in this work.  The sizes of the symbols correlate with the number of spectroscopic members.
  }
  \label{fig:m200_z}
\end{figure}

The first criterion is about groups themselves.  We want to 
reduce spurious groups that contaminate the analysis.
For this, we use groups with flag 1 or 2.
As mentioned above, they are spectroscopically confirmed groups.
We then restrict the sample to $0.5<z<1.0$ as shown in Fig. \ref{fig:m200_z}.
This redshift selection is motivated to have the {\sc [oii]} emission in our spectral coverage.
The line is not available at $z<0.5$ and we could in principle use H$\alpha$ to
fill that redshift range.  But, these two lines have different 
sensitivities to star formation and AGN
(e.g., H$\alpha$ is much more robust to extinction) and
it is hard to make fair comparisons between 
{\sc [oii]} emitters and H$\alpha$ emitters.
We do not include the $z<0.5$ groups in order to perform a robust analysis.
We do not include $z>1$ groups either as extremely strong fringes
in the VIMOS spectra decrease
the success rate of obtaining redshifts of passive galaxies at $z>1$, making
any spectral analysis at $z>1$ difficult.

In addition to the redshift criterion, we impose a group mass threshold.
As shown by previous studies (e.g., \cite{tanaka05,poggianti06,koyama07}),
galaxy properties depend on group mass.
In order to eliminate any strong group mass dependence and extract evolutionary trends,
we use groups with masses ($M_{200}$, which is the mass contained within $R_{200}$)
between $3\times10^{13}\rm M_\odot$ and
$7\times10^{13}\rm M_\odot$ as shown in Figure \ref{fig:m200_z}.
Note that $M_{200}$ is derived from the X-ray scaling relation calibrated against
weak lensing mass estimates \citep{leauthaud10}.
We still have a weak mass tendency within the narrow mass range in the sense that
we tend to have more massive systems at higher redshifts.
This bias {\it weakens} any evolutionary trends because more massive groups tend to
be more evolved.  If we had a flat group mass distribution at $0.5<z<1.0$,
we would have observed stronger evolutionary trends than those shown below.

The second criterion is the definition of group members.
In this work, we use galaxies with highly confident spectroscopic
redshifts from zCOSMOS.  Specifically, we use galaxies with flags
4's, 3's (including 14's and 13's), 2.5, 2.4, 9.5, 9.4, and 9.3.
For details of the flags, readers are referred to \citet{lilly09}.

To define the group membership, we first estimate $M_{200}$ from the X-rays.
From this, we evaluate the velocity dispersions ($\sigma$) and virial radii
($R_{200}$) of the groups assuming that they are virialized \citep{carlberg97}.
As most group galaxies are observed to lie within $\sim R_{200}$
in the local universe \citep{gomez03,tanaka04}, we define group members as
those within $<R_{200}$ from the centers.
The centers of groups with {\sc Flag}=2 can be uncertain, but this is not
a major concern in our analysis because a typical $R_{200}$ of these groups
is 2.5 times larger than the maximum positional uncertainty of 30 arcsec.
We have checked the robustness of our results by perturbing the centers of
the {\sc Flag}=2 groups with a Gaussian function with $\sigma=30$ arcsec
and repeated all the analysis in this paper.
We have observed no appreciable changes in our results.
For the line-of-sight separation, we apply $<2\sigma$ from the redshift centers
of the groups. Galaxies that do not belong to
any groups are defined as field galaxies.
In total, we have 7,549 galaxies with high confidence redshifts at $0.5<z<1$,
of which 246 are in groups satisfying the criteria above.
We will further apply a stellar mass cut to the galaxies in the next
section, and the total numbers of galaxies used in the main analyses
are 1,574 and 96 for the field and group environments, respectively.

%---------------------
%---------------------

%-------------------------------------------------------------------------------
\section{Galaxy Populations in the X-ray Groups at $0.5<z<1.0$}

%-------------------------------------------------------------------------------
\subsection{Color-mass diagram and the fraction of red galaxies}

%---------------------
\begin{figure*}
  \begin{center}
    \FigureFile(120mm,1mm){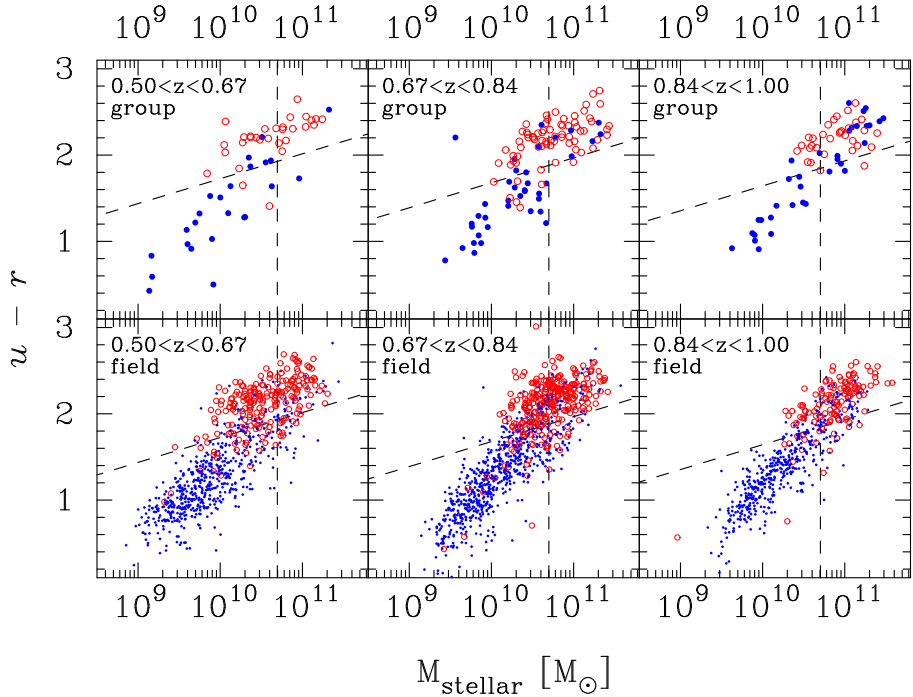}
  \end{center}
  \caption{
    Rest-frame $u-r$ color plotted against stellar mass.
    The top panels show the group galaxies and the bottom panels show the field galaxies.
    The panels are split in three redshift bins.
    The filled and open symbols show galaxies with and without significant {\sc [oii]} emission
    (EW{\sc [oii]}$<-5\rm\AA$ at $>2.3\sigma$), respectively.
    The vertical dashed lines are our mass threshold, which defines the stellar mass limited galaxy sample.
    We separate red and blue galaxies using the slanted dashed line.
    For clarify, we plot only one third of the field galaxies.
  }
  \label{fig:cmd}
\end{figure*}
%---------------------

We base our analysis on the group catalog and member catalog constructed
in the last section and study properties of galaxies as functions of
redshift and environment (i.e., group vs. field).  
To give an overview of the properties of galaxies in our catalog, 
we show a rest-frame $u-r$ color vs. stellar mass diagram in Fig. \ref{fig:cmd}.
The rest-frame color and stellar mass are derived by \citet{bolzonella10}
by fitting the photometry with model templates from \citet{bruzual03}
assuming the Chabrier initial mass function \citep{chabrier03}.
A typical error in our stellar mass estimates is $\sim0.2$ dex
(see \cite{bolzonella10} for details).
Before we discuss the plot, let us introduce our definition of
(a) red/blue galaxies and (b) {\sc [oii]} emitters.

{\bf (a)}
We divide galaxies into red and blue galaxies by their rest-frame $u-r$ color.
We perform a biweight fit to the red sequence in groups at $0.67<z<0.85$,
where we observe the most prominent red sequence due to the largest number of
group galaxies we have there.  We then shift it by $\Delta (u-r)=-0.3$ shown as
the slanted dashed line in Fig. \ref{fig:cmd} to separate the two populations.
The amount of the shift is motivated to give a reasonable separation between {\sc [oii]}
emitters and non-{\sc [oii]} emitters (see below for their definitions).
We have confirmed that our conclusions are insensitive to a small change in
the amount of the shift.
We correct for the color evolution of the red sequence in the other bins
using an instantaneous burst model formed at $z_f=3$ from \citet{bruzual03}.
As seen in the figure, the color threshold is bluer at higher redshift.

{\bf (b)}
We also use {\sc [oii]} emission to characterize galaxy properties.
We use the equivalent widths of {\sc [oii]} measured by Lamareille et al. (in prep).
Line detections below $1.15\sigma$ are considered fake and here we adopt
a conservative significance threshold of $2.3\sigma$ to ensure
that the line is securely detected.  We define galaxies with EW{\sc [OII]}$<-5\rm\AA$
detected at $>2.3\sigma$ as {\sc [oii]} emitters, and the other galaxies as quiescent.
Note we use a negative sign for emission.

Going back to Fig. \ref{fig:cmd}, it is immediately clear that the most massive galaxies
tend to be red and many of them do not show a sign of active star formation.
Galaxies with no significant {\sc [oii]} form a clear sequence of red galaxies.
Interestingly, some of the red galaxies show significant {\sc [oii]} 
emissions despite their red colors.  
In contrast to massive galaxies, low-mass galaxies are predominantly blue
{\sc [oii]} emitters.
This is due to a bias introduced by the flux limit of the survey.
The $I$-band, with which spectroscopic targets are selected
in zCOSMOS, samples bluer light in rest-frame at higher redshifts,
and we are missing low-mass red galaxies, resulting in a strong bias towards blue,
star forming galaxies.

It is hard to interpret Fig. \ref{fig:cmd} due to the strong selection bias.
We apply a stellar mass cut in each redshift bin to construct a stellar mass
limited sample in order to study evolutionary trends.
We have to correct for the mass evolution in each redshift bin, but
we cannot track the mass evolution of individual galaxies,
which depends on their star formation and merger histories.
Here, we simply apply correction for the passive evolution using
the same passive evolution model as used for the color evolution.
We can reach $\sim5\times10^{10}\rm  M_\odot$ galaxies
at $z=1$ in zCOSMOS, although the redshift success rate drops to
70\% (see Figs 2 and 10 of \cite{lilly09}).
\citet{cucciati10} and \citet{iovino10} applied a conservative cut of
$10^{11}\rm M_\odot$, but for the purpose of the paper, we do not need to
be 100\% complete and we apply a stellar
mass thresholds of $4.95\times10^{10}\rm M_\odot$,
$5\times10^{10}\rm M_\odot$, and $5.05\times10^{10}\rm M_\odot$ at
$0.5<z<0.67$, $0.67<z<0.84$, and $0.84<z<1$, respectively.
We note that our conclusions do not change if we adopt a conservative
mass cut of $10^{11}\rm M_\odot$, although the statistics becomes poor.
\citet{balogh11} reported on an abundant population of green valley galaxies
in groups at $0.85<z<1$.   We do not observe strong evidence for
an increased amount of green galaxies in Fig. \ref{fig:cmd}, but
we cannot probe as low-mass galaxies as they did ($10^{10.1}\rm M_\odot$).

%---------------------
\begin{figure}
  \begin{center}
    \FigureFile(80mm,1mm){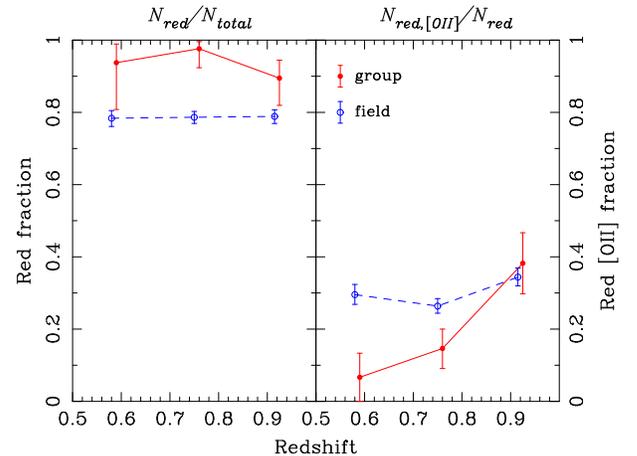}
  \end{center}
  \caption{
    {\bf Left:} Fraction of red galaxies plotted against redshift.
    Here we use the stellar mass limited sample in both the group and field environments.
    The filled and open symbols are for the group and field samples, respectively.
    They are slightly shifted horizontally to avoid overlapping.
    The error bars show 68\% confidence interval \citep{gehrels86}.
    {\bf Right:} Fraction of {\sc [oii]} emitters among red galaxies against redshift.
  }
  \label{fig:fred}
\end{figure}
%---------------------

We plot the fraction of red galaxies as a function of redshift in Fig. \ref{fig:fred}
using the stellar mass limited sample.
The fraction of massive red galaxies remains constant over the explored
redshift range in both groups and field.
An interesting trend in Fig. \ref{fig:fred} is that
the red fraction is always higher in groups than in the field, showing
clear environmental dependence of galaxy colors at $0.5<z<1$.
This may appear inconsistent with previous studies (e.g., \cite{cucciati10,iovino10}),
and we will discuss that in Section 4.1.
While we do not see strong color evolution,
we see a clear increase in
the fraction of {\sc [oii]} emitters among red galaxies at high redshifts
as shown in the right panel of Fig. \ref{fig:fred}.
As discussed earlier, our group sample is biased towards more massive groups
at higher redshifts, which weakens any evolutionary trends.  Also, due to
the nature of a mass limited sample, spectra at higher redshifts
are of lower signal-to-noise ratios.  We tend to miss weak {\sc [oii]}
emissions at higher redshifts, which weakens the trend we see here\footnote{
One might suspect that we tend to miss quiescent red galaxies at $z\sim1$
due to the busy OH lines, and
that might be driving the observed increase of red galaxies with
{\sc [oii]} emission, which are easier to identify.
However, as will be discussed in Section 3.2, red {\sc [oii]} emitters
increases with increasing stellar mass in the highest redshift bin.
This is an opposite trend from what expected from the redshift determination bias
because we do not miss massive (bright) passive galaxies even under the presence
of busy OH lines.
Therefore, the observed trend is unlikely due to an observational bias.
}.
The real evolutionary trend should be stronger than that observed
in Fig. \ref{fig:fred}.

The fraction of red {\sc [oii]} emitters in the field 
remains nearly constant with redshift, while it evolves very fast in groups.
The fractions become indistinguishable in between groups and field
in the highest redshift bin.
These red {\sc [oii]} emitters must have dusty star formation
activities and/or AGN activities.
This is a sort of evolution that cannot be revealed by broad-band photometry,
demonstrating a power of large spectroscopic surveys.

%-------------------------------------------------------------------------------
\subsection{Stellar mass and redshift dependence of the {\sc [oii]} emitters}

%---------------------
\begin{figure*}
  \begin{center}
    \FigureFile(120mm,1mm){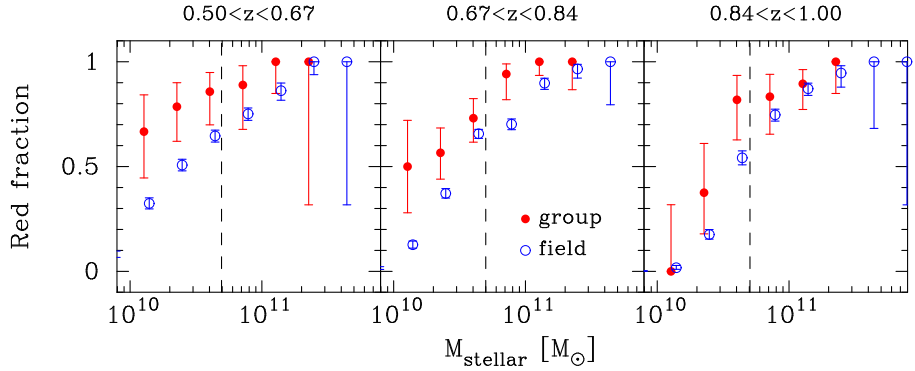}
  \end{center}
  \caption{
    Fraction of red galaxies as a function of stellar mass
    in the three redshift bins.
    The filled and open circles are the group and field galaxies, respectively.
    The vertical dashed lines are the mass thresholds, which define
    the stellar mass limited sample.
    The error bards are the 68\% confidence intervals.
  }
  \label{fig:smass_func3}
\end{figure*}
%---------------------

Essentially all galaxy properties are correlated with stellar mass.
It would be important to show that the higher red fraction in
groups observed in Fig. \ref{fig:fred}
is not due to the environmental dependence of the stellar mass function
such that groups host a larger fraction of massive galaxies,
which increases the red fraction because massive galaxies tend to be red.
We plot in Fig. \ref{fig:smass_func3} the fraction of red galaxies
as functions of stellar mass and redshift.  
We find that the red fraction tends to be higher in groups than in the field
at a given stellar mass, although the error bars often overlap.
The difference between groups and field is particularly clear at $0.67<z<0.84$,
where we have a prominent large-scale structure \citep{guzzo07}.
The poor statistics does not allow us to conclude that the groups show a
higher red fraction at a given stellar mass, but the systematically higher
fraction suggests that the observed high red fraction in groups
in Fig. \ref{fig:fred} is not entirely due to the dependence of stellar mass
function on environment.
We note that \citet{george11} showed that the red fraction is
higher in groups than in the field at a given stellar mass based on
the same X-ray group catalog and on the photometric data in COSMOS.

%---------------------
\begin{figure*}
  \begin{center}
    \FigureFile(120mm,100mm){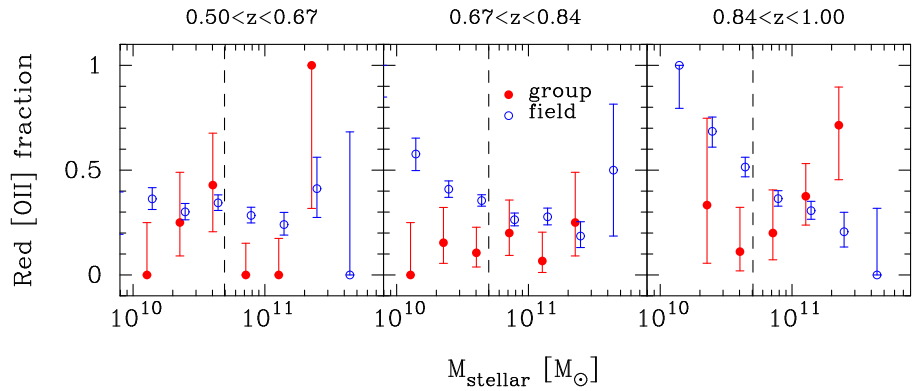}
  \end{center}
  \caption{
    Fraction of red {\sc [oii]} emitters as a function of stellar mass.
    As in Fig. \ref{fig:smass_func3}, the panels are split into the three redshift bins.
    The vertical dashed lines are the stellar mass cuts.
    The error bars are the 68\% confidence intervals.
  }
  \label{fig:smass_func2}
\end{figure*}
%---------------------

Similarly, it would be interesting to look at the stellar mass
dependence of the {\sc [oii]} emitters on the red sequence.
We plot in Fig \ref{fig:smass_func2} the fraction of the red {\sc [oii]}
as a function of stellar mass.
In the field, the fraction of the red {\sc [oii]} emitters
does not strongly depend on mass above the mass threshold.
The overall fraction does not show a strong increase with redshift either.
On the other hand, the fraction in groups increases
at higher redshift, and in the highest redshift bin, 
the fraction of {\sc [oii]} emitters seems to increase with increasing
mass above the mass cut.  Although the statistics is poor, this
is a contrasting trend to the field.
It seems that the increase of {\sc  [oii]} emitters
in groups at high redshift  is stronger for more massive galaxies.

A high fraction of {\sc [oii]} emitters at $z\sim1$ 
has already been observed by several authors.  
\citet{nakata05} found a fraction of EW{\sc [oii]}$<-10\rm \AA$ galaxies
of $\sim0.45\pm0.15$ at $0.8<z<1$.  Their sample is not
stellar mass limited and we cannot make a fair comparison with theirs.
But, if we apply the same selection of EW{\sc [oii]}$<-10\rm \AA$
to our sample, we obtain a consistent fraction of $0.31\pm0.07$.
\citet{poggianti06} showed the high fraction (very roughly 50\%)
in groups and clusters at $0.4<z<0.8$.
If we apply EW{\sc [oii]}$<-3\rm \AA$ as done in \citet{poggianti06},
we obtain a consistent fraction.
A similarly high fraction of {\sc [oii]} emitters in groups
at higher redshift ($z=1.2$) is reported by \citet{tanaka09}.
If we apply our definition of {\sc [oii]} emitters and
the stellar mass cut to their sample,
we find that the fraction of massive {\sc [oii]} emitters is
$0.47\pm0.22$.
We should be careful with this face value as their sample was drawn
from optically selected groups and the targets for spectroscopy
were photo-$z$ selected, which potentially introduces biases
in the sample.  However, this high fraction of {\sc [oii]}
emitting galaxies in groups drawn from completely different
sample is reassuring.
Unfortunately, their sample is not large enough to constrain
the mass dependence of the {\sc [oii]} emitters.

\begin{figure}
  \begin{center}
    \FigureFile(80mm,1mm){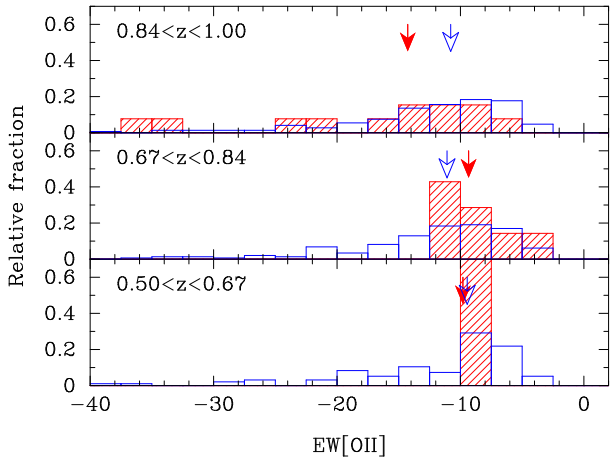}
  \end{center}
  \caption{
    Distribution of EW{\sc [oii]} of red galaxies measured at $>2.3\sigma$
    in groups (shaded histogram) and in the field (open histogram)
    environments based on the stellar mass limited sample.
    The arrows point the median of the EW{\sc [oii]} in each environment.
  }
  \label{fig:ewoii_distrib}
\end{figure}

Not only the fraction of red {\sc [oii]} emitters, but
strengths of the emission seem to increase with increasing
redshift as shown in Fig. \ref{fig:ewoii_distrib}.
Field galaxies do not show any strong change in the median 
EW{\sc [oii]} with redshift.
On the other hand, group galaxies seem to show a larger EW{\sc [oii]} tail
at higher redshifts.
The distributions of EW{\sc [oii]} in groups at $0.67<z<0.84$
and $0.84<z<1$ show a null probability of 3\%
from the Mann-Whitney $U$ test.
This increase in EW{\sc [oii]} would not be too surprising
given the rapid increase in the fraction of {\sc [oii]} emitters observed
in Fig. \ref{fig:fred}, which is statistically significant.
We have only one red {\sc [oii]} emitter at $0.50<z<0.64$,
and we cannot apply any statistical tests there.
In contrast to groups, field galaxies do not show any strong evolutionary trends:
the Mann-Whitney test gives a null probability of 33\% between
$0.50<z<0.64$ and $0.67<z<0.84$, and 47\% between $0.67<z<0.84$ and $0.84<z<1$.
\citet{wilman08} showed that $\sim50\%$ of galaxies with $10^{11}\rm M_\odot$
in $z\sim0.4$ groups show infrared flux excess, which can be due to
star formation and/or AGN.  The observed low frequency of the {\sc [oii]} emitters
compared to the infrared detections may
be because {\sc [oii]} and infrared have different sensitivities to
star formation and AGN.  It could also be because of our conservative EW{\sc [oii]} cut.

To summarize, we observe that the fraction of red galaxies with $>5\times10^{10}\rm M_\odot$
does not strongly evolve at $0.5<z<1$ in both group and field environments,
and it is always higher in groups than in the field.
The most striking trend that we find is that
the fraction of red {\sc [oii]} emitters in groups increases at higher redshifts,
while the fraction is nearly constant in the field.
It seems that more massive galaxies in groups show stronger increase in {\sc [oii]}.
This trend suggests that the red galaxies in groups have dusty star formation
and/or AGN activities and the rates at which environment suppresses such activities  are
different in different environments.
We will pursue this point in the next section.

%---------------------
%---------------------

%-------------------------------------------------------------------------------
\section{Discussions}

%---------------------
\subsection{Comparisons with previous studies}

Recent large spectroscopic surveys such as zCOSMOS \citep{lilly07},
DEEP2 \citep{davis03}, and VVDS \citep{lefevre05} have enabled statistical
analysis of galaxies in the universe up to $z=1$ and even beyond.
Several authors have studied the environmental dependence of galaxy
properties using data from those surveys
(e.g., \cite{cucciati06,cooper07,gerke07,cooper10,cucciati10,iovino10}).
But, the results from those papers are not always consistent.
In particular, results on the color-density relation at $z\sim1$
and interpretations of it seem to be controversial.

In fact, our finding that the fraction of red galaxies depends
on environment up to $z\sim1$ does not seem to be consistent at a first glance
with those from \citet{cucciati10} and \citet{iovino10} who have found
no strong dependence of galaxy colors on density at $z\sim1$
based on a stellar mass limited sample.
Another finding by \citet{iovino10} that the red fraction decreases with increasing
redshift is not consistent either.
\citet{cooper10} discussed differences in the data sets and in the ways
analyses were made in the literature.
But, we use basically the same data set as \citet{cucciati10} and
\citet{iovino10} and the observed differences appear at odds.
Here we attribute the cause of the differences to differences in (i)
definitions of environments and in (ii) definitions of red galaxies.

There is a fair amount of evidence that galaxy properties
depend on the mass of groups and clusters at high redshifts
(e.g., \cite{tanaka05,poggianti06,koyama07}).
In this paper, we use X-ray selected groups, while most of the previous
papers from large spectroscopic surveys are based on galaxy groups
(or galaxy densities) identified using spectroscopically observed galaxies.
There are pros and cons in these environment definitions, but
the advantage of the X-ray groups is that our environment is mass-selected.
As mentioned earlier, 90\% of the groups at $z<1$ in COSMOS are
spectroscopically confirmed, and we probably sample the group-mass
environments well in this study.  A further comparison between
optical and X-ray groups will be made in Finoguenov et al. (in prep).

A disadvantage would be that we cannot identify very low-mass
groups which are below our X-ray detection limit, while optical
group identification algorithms can.  We suspect that these
low-mass groups would be the primary cause of
the difference from \citet{cucciati10} and \citet{iovino10}.
\citet{iovino10} showed that optically poor groups tend to
exhibit a lower red fraction than optically rich groups.  Poor groups
are more numerous than rich ones and they dominate the group sample, which
results in a smaller difference in the red fraction between
groups and field.
The difference in the group mass ranges explored could account for
the difference in the dependence of the red fraction on environment
between us and the previous studies.
We note that \citet{george11} recently observed clear dependence of
the red fraction as a function of stellar mass on environment based on
the photometric data in COSMOS.

Another cause of the difference is that our definition
of red galaxies is different from that adopted in \citet{iovino10}.
They adopted a color threshold of $U-B=1$ regardless of redshift
and stellar mass of galaxies, while we account for the tilt of
the red sequence with respect to stellar mass and also for the
passive evolution.  We have confirmed that the red fraction
decreases at higher redshifts if we adopt the same definition as theirs.
We prefer to account for the tilt and passive evolution to define
red/ blue galaxies in this paper.
The above two reasons are likely the primary causes of the somewhat
different results between us and the previous authors.

\citet{gerke07} and \citet{cooper07} also suggested that
the red fraction decreases with increasing redshifts in high density environments.
This might be due to the selection of galaxies.
They applied a rest-frame $B$-band magnitude cut, not a stellar mass cut.
The $B$-band magnitude cut introduces a strong bias towards star forming galaxies,
which could result in a lower red fraction at higher redshifts.

We note that there is a non-zero possibility that we are missing
groups dominated by blue galaxies because we used the red sequence
finder in the group identification process and that could possibly enhance
the difference between groups and field.
However, our technique uses a contrast of the red sequence
between group and field and we do not actually require a higher fraction
of red galaxies in groups.
Even if groups had the same red fraction as the field, 
we can identify them as long as they show an over-density.
We miss only groups in which the fraction of red galaxies is
significantly lower than the field.
It is unlikely that such very blue groups are so abundant that
they change our results significantly because we have identified
$\sim90\%$ of all the X-ray group candidates at $z<1$ with high significance
and they exhibit a clearer red sequence compared to field galaxies.

Recently, \citet{koyama11} reported on H$\alpha$ narrow-band imaging of a $z=0.4$ cluster.
They found that groups exhibit a higher fraction of H$\alpha$ emitters
than in the field, which is in contrast to our finding in the right panel of Fig. \ref{fig:fred}.
There are a number of differences in the explored stellar mass range,
observing technique, emission line used and emission line sensitivity.
These differences hinder detailed comparisons with our results.

%---------------------
\subsection{Origin of the {\sc [oii]} emission -- star formation vs AGN}

The increasing fraction of red {\sc [oii]} emitters with increasing redshift
must be due to increasing dusty star formation activities and/or AGN activities.
Recent studies of $z>1$ clusters also reported on increased rate of
emission line galaxies in clusters than in lower redshift clusters
(e.g., \cite{hayashi10}, but see also \cite{bauer10}).
But, this trend is not established yet, and it is not clear
if the emission line originates from star formation or
AGN activities either.

%---------------------
\begin{figure}
  \begin{center}
    \FigureFile(80mm,1mm){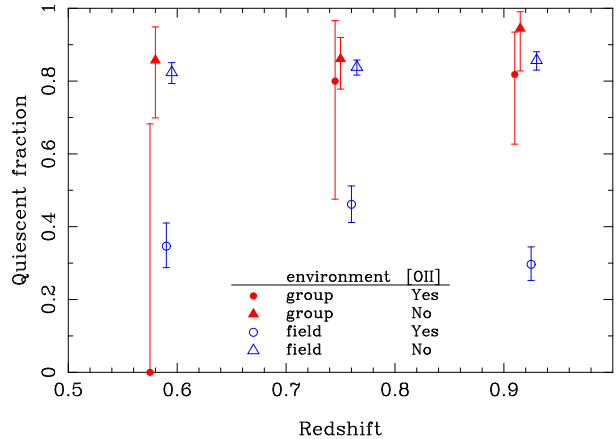}
  \end{center}
  \caption{
    Fraction of quiescent galaxies with $(NUV-r)_{dered}>3.5$ among
    the red galaxies in groups and field with/without {\sc [oii]}
    plotted against redshift.
    The meanings of the symbols are shown in the plot.
    The error bars show the 68\% confidence intervals.
  }
  \label{fig:nuv_r_distrib2}
\end{figure}
%---------------------

Let us first ask if the {\sc [oii]} emission is due to dusty star formation.
We look at the $(NUV-r)_{dered}$ color of the {\sc [oii]} emitters taken
from \citet{ilbert09}.  The $(NUV-r)_{dered}$ color is a reddening
corrected color of galaxy templates used for the photometric redshift
estimates (i.e., the raw template colors without dust extinction)
using the 30-band photometry.
This is sensitive to on-going star formation as shown by \citet{ilbert10}.
We adopt a threshold of $(NUV-r)_{dered}=3.5$
to separate quiescent galaxies from star-forming galaxies \citep{ilbert10}
and plot the fraction of quiescent galaxies 
in Fig. \ref{fig:nuv_r_distrib2}.  As the color from \citet{ilbert09}
is based on photometric redshifts, we use galaxies with correct 
photometric redshifts ($|z_{phot}-z_{spec}|<0.05$).
We note that this analysis is essentially equivalent to the popular
two-color diagnostics to separate quiescent red galaxies from dusty ones
\citep{wolf05,wolf09}.
But, instead of using only 2 colors, we here make use of the 30-band
photometry that gives a fine sampling of galaxy SEDs over a wide wavelength
range to discriminate quiescent galaxies from dusty star forming ones.

Let us start with the red galaxies without significant {\sc [oii]} emission
shown by the open and filled triangles.
The quiescent fractions of these galaxies are very high
as expected from the absence of {\sc [oii]}.
The stellar population of these galaxies is typically old and there is
no significant difference between groups and field.
That is, if red galaxies exhibit no {\sc [oii]}, they are dominated
by old stellar populations regardless of environment.
Now, we turn our attention to the red {\sc [oii]} emitters.
The quiescent fraction of the red  {\sc [oii]} emitters in the field
(open circle) is relatively low ($30-40\%$),
suggesting that more than a half of them are not quiescent and are probably
undergoing dusty star formation.
In contrast, the red {\sc [oii]} emitters in groups (filled circle)
show a very high quiescent fraction even at high redshifts
and it is as high as those without {\sc [oii]} emission.
This suggests that red {\sc [oii]} emitters in groups are dominated by
old stellar populations despite the {\sc [oii]} emission.
In Fig. \ref{fig:fred}, we observed a sharp increase in the fraction of
red {\sc [oii]} emitters in high redshift groups, but
the quiescent fraction does not show a corresponding decrease.
There is no clear evidence for increased dusty galaxies in the red {\sc [oii]}
emitters in groups.
Instead, Fig. \ref{fig:nuv_r_distrib2} favors the AGN origin.

Let us then stand on the other side of the view and ask whether
the {\sc [oii]} emission comes from AGNs.
At the redshift range under study, we cannot use strong emission line
diagnostics such as the one proposed by \citet{baldwin81} to identify
AGNs as the H$\alpha$ line migrates to near-IR.
Here we take another way to identify AGNs -- X-rays -- and quantify
how AGNs populate in the redshift range studied here.
We use the Chandra point source catalog \citep{elvis09}
and apply an X-ray luminosity cut of $L_{0.5-10keV}>10^{43}\rm\ erg\ s^{-1}$,
at which we are nearly complete up to $z=1$.
In the field, an X-ray detection rate of the red {\sc [oii]} emitters
seems to increase in the highest redshift bin:
$0.04\pm0.03$, $ 0.03\pm0.02$, and $0.13\pm0.04$ from low to high redshift bins.
However, we detect no red {\sc [oii]} emitters in groups in X-rays:
$0.00^{+0.68}_{-0.00}$, $0.00^{+0.32}_{-0.00}$, and $0.00^{+0.17}_{-0.00}$
from low to high redshifts.

Most of the red {\sc [oii]} emitters in groups are not detected in X-rays
(only 2 are detected, but with luminosities below the cut applied above).
\citet{tanaka11} found that AGNs in quiescent galaxies are typically
soft, low-luminosity AGNs in the local universe. Motivated by this,
we performed a stacking analysis of those undetected sources in the soft band
with a special care to remove the extended component \citep{finoguenov09}.
By stacking 9 objects that are not individually detected,
we measure an average luminosity of $2.8\times10^{-17}\rm\ erg\ s^{-1}\ cm^{-2}$
in 0.5--2 keV, which translates into $2.7\times10^{40}$ and
$1.5\times10^{41}\rm\ erg\ s^{-1}$ at $z=0.5$ and $z=1$, respectively.
This luminosity level can be explained both by low-luminosity AGN
and star formation origins.
We cannot constrain the AGN fraction in groups with X-rays.

X-rays unfortunately do not put any constraint on the AGN
or dusty star formation origin, but the robust photometric analysis
based on the 30 photometric bands presented above seems to lend support
to the AGN origin of the {\sc [oii]} emission, although the statistics is poor.
The stellar population of the red {\sc [oii]} emitters in groups is old
and there is no hint of strong on-going star formation in those galaxies.
The observed {\sc [oii]} emission is unlikely due to star formation
and the most probable origin of it is AGNs.

A possibility of weak AGNs in groups is supported by recent work by \citet{lemaux10},
who performed near-IR spectroscopy of galaxies dominated by old stellar
population but have {\sc [oii]} emission in $z=0.8$ and 0.9 clusters.
They found that a significant fraction of them ($\sim70\%$) harbor AGNs.
It would not be surprising if a large fraction of the red {\sc [oii]}
emitters in our high redshift groups are AGNs as they have similar photometric properties
as those studied in \citet{lemaux10}.
However, other authors reported on increased dusty star formation activities
in groups at high redshifts (e.g., \cite{koyama08,koyama10,kocevski10}).
\citet{tanaka09} found that group galaxies at $z\sim1.2$
show weak H$\delta$ absorptions and they speculated that it might be
due to large extinction.
Post-starburst galaxies might favor groups \citep{poggianti09}.
\citet{vergani10} also reported that post-starburst galaxies prefer
high density environments based on the zCOSMOS data.
Recently, \citet{hayashi11} observed that both AGN and star formation
take place in a $z=1.4$ cluster.

Given this controversial situation, it is probably fair to say that
the origin of the emission line is still unclear at this point.
It may be that both dusty star
formation activities and AGN activities increases at high redshifts
and there is a strong cluster-cluster variation.
Any conclusion on the origin of the {\sc [oii]} emission needs to be
drawn from a larger statistical sample of groups at $z\gtrsim1$.
An extensive near-IR spectroscopy targeting H$\alpha$ and {\sc [nii]}
lines of the red {\sc [oii]} emitters in groups to
perform emission line diagnostics such as the one proposed by
\citet{baldwin81} would be an obvious way forward.
Also, the newly developed AGN identification method by \citet{tanaka11} is
effective as well because it requires only {\sc [oii]} and/or {\sc [oiii]}.
Deep Chandra observations are obviously helpful as well.
Using these techniques, we first have to discriminate AGNs from
star formation in order to interpret the recent observations that
distant groups and clusters tend to show an increased rate of emission line galaxies.

%-------------------------------------------------------------------------------
\section{Summary}

We have presented photometric and spectroscopic analyses of zCOSMOS galaxies
at $0.5<z<1$.  Unlike most of the previous studies, we define the mass-selected
environments to study the dependence of galaxy properties on environment.
This is the most important feature of this work.  Previous studies have shown
that galaxy properties depend on mass of groups and clusters.
These studies clearly show that environment needs to be defined by mass.

We have found that the fraction of red galaxies is always higher
in groups than in the field at $0.5<z<1$ and it does not strongly
change over this redshift range.
This result might appear
inconsistent with previous studies from zCOSMOS, but we have argued
that this is due to different environment definitions and to different
definitions of red galaxies.
The most important finding of this paper is that the fraction of
{\sc [oii]} emitters on the red sequence increases in groups
at higher redshifts, while the fraction does now show any significant
evolution in the field.  The increased red {\sc [oii]} emitters in groups must be due
to increased dusty star formation activities and/or to increased
AGN activities.  
We have studied these two possibilities by using the 30-band photometry
and X-ray data.  While the X-ray data do not put a strong constraint on
them, the 30-band photometry suggests that the stellar population of the
{\sc [oii]} emitters in groups is old and there is no hint of
enhanced dusty star forming activities.  This lends support to
increased AGN activities.

Recent observations often report a high fraction of emission line
galaxies in distant groups and clusters.
The question now is where the emission comes from.
We have obtained evidence for the AGN origin and recent
near-IR spectroscopic work also favors it.  But, our overall statistics
is poor and some of the previous studies seem to favor the dusty star formation origin.
More observations are obviously needed to settle the issue.

\vspace{0.5cm}

This work is supported by World Premier International Research Center Initiative
(WPI Initiative), MEXT, Japan and also in part by KAKENHI No. 23740144.
This work is based on observations undertaken at
the European Southern Observatory (ESO) Very Large Telescope (VLT) under
the Large Program 175.A-0839.
We would like to thank the anonymous referee for useful comments,
which helped improve the paper.

%-------------------------------------------------------------------------------
%-------------------------------------------------------------------------------

\end{document}